\documentclass[aps,nofootinbib,twocolumn,prl,showpacs,class-pre]{revtex4}
\usepackage{setspace} \usepackage{amsmath,amssymb,amsfonts,amsthm}
\usepackage{graphicx}

\newcommand{\be}{\begin{equation}}
\newcommand{\ee}{\end{equation}}
\newcommand{\beq}{\begin{eqnarray}}
\newcommand{\eeq}{\end{eqnarray}}

\begin{document}
\title{Cosmology and the Noncommutative approach to the Standard Model}
\author{William Nelson\footnote{nelson@gravity.psu.edu}}
\affiliation{ Institute of Gravitation and the Cosmos, Penn State
University, State College, PA 16801, U.S.A.}
\author{Mairi Sakellariadou\footnote{mairi.sakellariadou@kcl.ac.uk}}
\affiliation{Department of
  Physics, King's College, University of London, Strand WC2R 2LS,
  London, U.K.}

\begin{abstract}
We study cosmological consequences of the noncommutative approach to
the standard model of particle physics. Neglecting the nonminimal
coupling of the Higgs field to the curvature, noncommutative
corrections to Einstein's equations are present only for inhomogeneous
and anisotropic space-times. Considering the nonminimal coupling
however, corrections are obtained even for background cosmologies.
Links with dilatonic gravity as well as chameleon cosmology are
briefly discussed, and potential experimental consequences are
mentioned.
\end{abstract}


\maketitle

\section{Introduction}
Theoretical early universe cosmology is gaining a constantly
increasing interest from the scientific community. The predictions of
the theoretical models can now be compared with a plethora of
astrophysical data, in particular the measurements of the Cosmic
Microwave Background temperature anisotropies; all having a surprising
good accuracy. Moreover, present high energy experiments, in
particular the Large Hadron Collider, will test some of the
theoretical pillars of the cosmological models. Despite this golden
era of cosmology, a number of questions, such as the explanation of
space-time dimensionality~\cite{dim}, the origin of dark
energy~\cite{de} and dark matter~\cite{bergstrom}, the search for the
natural and well-motivated successful inflationary model, are still
awaiting for a definite answer.

The main theoretical approaches upon which the cosmological models
have been built are either string theory or quantum gravity. Here we
will consider another one, which up to now has, rather surprisingly,
gained only a limited interest, namely NonCommutative Geometry
(NCG)~\cite{NCG1,NCG2}. More precisely, we will study cosmological
consequences of the NCG approach to the Standard Model
(SM)~\cite{ccm}, which remains the best of our knowledge particle
physics model at present. The NCG approach leads to all the detailed
structure of the SM, as well as several physical predictions at the
unification scale.

In this paper, after a brief review on the noncommutative spectral
action, we discuss some of its early universe cosmological
consequences and their potential link to dilatonic gravity and
chameleon models.

\section{Noncommutative Geometry approach}
The NCG approach to the unification of all fundamental interactions
including gravity is based on three ansatz:

\medskip

(I) Slightly below Planck energy, space-time becomes the product of a
four-dimensional smooth compact Riemannian manifold $\mathcal{M}$ by a
finite noncommutative space $F$. The geometry is therefore the tensor
product of an internal geometry for the SM and a continuous geometry
for space-time. One has to distinguish between the metric (or
spectral) dimension, given by the behaviour of the eigenvalues of the
Dirac operator, and the KO-dimension, an algebraic dimension based on
K-theory. The relevant Dirac operator for space-time is the ordinary
Dirac operator on curved space-time, thus the metric dimension is
equal to four. The internal Dirac operator consists of the fermionic
mass matrix, which has a finite number of eigenvalues, thus the
internal metric dimension is zero.  As a result, the metric dimension
of the product geometry is four, the same as the ordinary space-time
manifold.

To resolve the fermion doubling problem, by projecting out the
unphysical degrees of freedom resting in the internal space, the real
structure of the finite geometry $F$ must be such that its
KO-dimension is equal to six~\cite{fdp}. Thus, the KO-dimension of the
product space $\mathcal{M}\times F$ is equal to $10\sim 2\ {\rm
  modulo}\ 8$.  Notice that unlike earlier particle physics models
based on NCG, in the approach~\cite{ccm} followed here the
KO-dimension (which is equal to 6 modulo 8) of the internal space is
different than its metric dimension (which is equal to zero).

A noncommutative geometry is given by a representation of spectral
nature. More precisely, $F=(\mathcal{A}, \mathcal{H}, D)$ is a
spectral triple, given by an involutive algebra $\mathcal{A}$ of
operators in Hilbert space ${\cal H}$, playing the r\^ole of the
algebra of coordinates, and a linear self-adjoint ($D=D^\dagger$)
operator $D$ in $\mathcal{H}$, playing the r\^ole of the inverse of
the line element.  The choice of Hilbert space ${\cal H}$ is
irrelevant here, since all separable infinite-dimensional Hilbert
spaces are isomorphic. The operator $D$ is such that all commutators
$[D,a]$ are bounded for $a\in \mathcal{A}$.  Except for finite
dimensional cases, $D$ is in general not a bounded operator, hence it
is only defined on a dense domain.

The classic geodesic formula of Riemannian geometry:
\be d(x,y)={\rm inf}\int_\gamma {\rm d} s~, 
\ee 
where the infimum is taken over all paths from $x$ to $y$, giving the
distance $d(x,y)$ between two points $x,y$, is replaced in NCG by
\be d(x,y)={\rm sup}\{|f(x)-f(y)|:f\in{\mathcal A},
||[\mathcal{D},f]||\leq 1\}~, 
\ee 
with $D$ the inverse of the line element ${\rm d}s$.  Another
significance of $D$ is that its homotopy class represents the
K-homology~\footnote{K-homology is the homological version of
  K-theory.} fundamental class of the space under consideration.

The choice of the finite dimensional involutive algebra consists of
the main input for the model.  The hypothesis that space-time is the
product of a continuous manifold $M$ by a discrete space $F$ is the
easiest generalisation of a commutative space. This is a strong
assumption that is expected to break in the Planck era.

\medskip

(II) The algebra constructed in this product space-time is then
assumed to be in the {\sl symplectic-unitary}
case~\cite{Chamseddine:2007ia}. This choice restricts the algebra
$\mathcal{A}$ to the form $\mathcal{A}=M_{a}(\mathbb{H})\oplus
M_{k}(\mathbb{C})$, with $k=2a$; $\mathbb{H}$ is the algebra of
quaternions. The first possible value for the even number $k$ is 2,
corresponding to a Hilbert space of four fermions; it is ruled out
from the existence of quarks. The second one, $k=4$, leads to the
correct number of $k^2=16$ fermions in each of the three
generations. Notice that considering three generations is a physical
input in NCG~\cite{Chamseddine:2007ia}. The involutive algebra
$\mathcal{A}$ corresponds to a given space in the same way as in the
classical duality between {\sl space} and {\sl algebra} in algebraic
geometry.

\medskip

(III) The Dirac operator connects the two pieces of the product
geometry nontrivially. The action $S$, called the {\sl spectral
  action} functional, depends only of the spectrum of the Dirac
operator; it is of the form ${\rm Tr}(f(D/\Lambda))$, with $\Lambda$
giving the energy scale and $f$ being a test function, whose choice
plays only a small r\^ole.  The spectral action functional ${\rm
  Tr}(f(D/\Lambda))$ accounts only for the bosonic term; the fermionic
term can be included by adding $(1/2)\langle J\psi,D\psi\rangle$.
When the spectral action $S$ is expanded in inverse powers of
$\Lambda$, it depends only on three first momenta $f_k=\int_0^\infty
f(v)v^{k-1}{\rm d}v$ for $k>0$, and on the Taylor expansion of $f$ at
0. One of the consequences is that some of the fermions can acquire
Majorana masses, realising the see-saw mechanism.

The full Lagrangian of the SM, minimally coupled to gravity, is
obtained~\cite{ccm} as the asymptotic expansion of the spectral action
for the product space-time. For our purposes here, namely extracting
early universe cosmological consequences of the noncommutative
spectral action approach, we are only interested in the gravitational
and Higgs part of the action, namely
\beq\label{eq:action1} {\cal S}_{\rm grav}^{\rm Lorentzian} = \int \left(
\frac{1}{2\kappa_0^2} R + {1\over 2}\alpha_0
C_{\mu\nu\rho\sigma}C^{\mu\nu\rho\sigma} + \tau_0 R^\star R^\star\right.
\nonumber\\
-\left.
\xi_0 R|{\bf H}|^2 \right) \sqrt{-g} {\rm d}^4 x~; \eeq
$ {\bf H} $ is a rescaling ${\bf H} = (\sqrt{a f_0}/\pi) \phi $ of the
Higgs field $\phi$ to normalise the kinetic energy. The momentum
$f_0=f(0)$, is physically related to the coupling constants at
unification.  The coefficient $a$, that enters the Higgs field
redefinition, is given by
\beq\label{eq:Ys}
 a&=&{\rm Tr} \left( Y^\star_{\left(\uparrow 1\right)}
Y_{\left(\uparrow 1\right)} +  Y^\star_{\left(\downarrow 1\right)}
Y_{\left(\downarrow 1\right)}\right.
\nonumber\\ 
&&+ \left. 3\left( 
 Y^\star_{\left(\uparrow 3\right)} Y_{\left(\uparrow 3\right)}
+  Y^\star_{\left(\downarrow 3\right)} Y_{\left(\downarrow 3\right)}
\right)\right)~,
\eeq
where the $Y$'s are used to classify the action of the Dirac operator
and give the fermion and lepton masses, as well as lepton mixing, in
this asymptotic version of the spectral action. The $Y$'s matrices are
only relevant for the coupling of the Higgs field with fermions
through the the dimensionless matrices $\pi/\sqrt {af_0} Y_x$ with
  $x\in \{(\uparrow\downarrow ,j)\}$. Thus, $a$ has the physical
  dimension of a (mass)$^2$.

The coupling constants in Eq.~(\ref{eq:action1}) are
\beq \label{coupl-const} \frac{1}{\kappa_0^2} = \frac{ 96 f_2
  \Lambda^2 - f_0 c^2}{12\pi^2}~,\nonumber\\ \alpha_0 =
-\frac{3f_0}{5\pi^2}~~, ~~\tau_0 = \frac{11f_0}{60\pi^2}~~,~~ \xi_0 =
\frac{1}{12}~, \eeq
where $\Lambda$ is an energy scale about which the asymptotic
expansion is performed and $c$ is expressed in terms of $Y_R$ which
gives the Majorana mass matrix, $c={\rm Tr} \left( Y^\star_R Y_R
\right)$.  The scale $\Lambda$ is fixed by the unification scale of
the coupling constants of the Standard Model.  Let us emphasise that
the spectral action, Eq.~(\ref{eq:action1}), has to be seen as a
boundary condition at unification scale. Therefore,
Eq.~(\ref{coupl-const}) above fixes the coupling constants at
unification scale; extrapolations to lower energies are possible using
renormalisation group analysis. It is therefore evident that this
noncommutative spectral action approach is appropriate for early
universe cosmology ({\sl i.e.}, at energies close to unification).

Several key points should be noted: Firstly, the noncommutative
geometry procedure outlined above is entirely classical; it simply
provides an elegant way in which the Standard Model of particle
physics can be produced from purely (noncommutative) geometric
information. Secondly, the action given in Eq.~(\ref{eq:action1}) has
been Wick rotated from the Euclidean action which is produced from
noncommutative geometry~\footnote{To use the formalism of spectral
  triples in noncommutative geometry, it is convenient to work with
  Euclidean rather that Lorentzian signature. One can go to Euclidean
  signature by performing a Wick rotation to imaginary time. In the
  Euclidean action functional for gravity, the kinetic terms must have
  the correct sign so that the functional is bounded below. Since such
  positivity is spoiled by the scalar Weyl mode, one must show that
  all other terms get a positive sign~\cite{NCG2}.}. The formal
justification of this has yet to be shown. Thirdly, at present the
entries in the Dirac operator that produce Eq.~(\ref{eq:Ys}) are
inputs to the theory. The hope is that by varying with respect to
them, the values that correspond to the Standard Model will be
dynamically chosen. Despite these issues, it remains striking that by
removing the assumption that space-time is commutative in the simplest
possible way (the space-time is a product of a commutative manifold
$M$ and a discrete, internal, noncommutative manifold $F$), one
recovers General Relativity coupled to the entire Standard Model with
no additional particles and the correct couplings.

The only nonstandard elements of the asymptotic expansion of the
noncommutative geometry action are the presence of the additional
terms given in Eq.~(\ref{eq:action1}).  The purpose of this paper is
indeed to investigate some cosmological consequences of these terms.

\section{Cosmological consequences}
Let us study the gravitational part of the spectral action
Eq.~(\ref{eq:action1}).  The first two terms give the Riemannian
curvature with a contribution from the Weyl curvature, where the
second term is the action for conformal
gravity~\cite{mannheim}. Notice that the presence of the
Einstein-Hilbert term (and of the cosmological constant, which we
neglect here) explicitly break conformal invariance.  The third term
is a topological term integrating to the Euler characteristic of the
manifold:
\be R^\star R^\star =\frac{1}{4} \epsilon^{\mu\nu\rho\sigma}
\epsilon_{\alpha\beta\gamma\delta} R^{\alpha\beta}_{\mu\nu}
R^{\gamma\delta}_{\rho\sigma} ~,\nonumber \ee
hence is nondynamical. Finally, the fourth term is the scalar mass
term.

The equations of motion arising from Eq.~(\ref{eq:action1})
read~\cite{mannheim}:
\beq\label{eq:EoM1} 
&&R^{\mu\nu} - \frac{1}{2}g^{\mu\nu} R -
\alpha_0\kappa_0^2\delta\left(\Lambda\right) \left[
  2C^{\mu\lambda\nu\kappa}_{\ \ \ \ \ \ ;\lambda ; \kappa} -
  C^{\mu\lambda\nu\kappa}R_{\lambda \kappa}\right]\nonumber\\ 
&&\ = \ 
 \kappa_0^2\delta\left(\Lambda\right)T^{\mu\nu}_{\rm matter}~, \eeq
where \be \delta\left( \Lambda\right) \equiv [1 -2\kappa_0^2
  \xi_0|{\bf H}|^2]^{-1}~. \nonumber \ee 
In what follows, we study the above equations of motion first
neglecting the nonminimal coupling between the geometry and the Higgs
field and then including it.

\subsection{Neglecting the Higgs field term}
Neglecting the nonminimal coupling between the geometry and the Higgs
field, {\sl i.e.}, setting $\phi = 0$ in Eq.~(\ref{eq:EoM1}), leads to:
\beq\label{eq:EoM2}
R^{\mu\nu} - \frac{1}{2}g^{\mu\nu} R - \alpha_0\kappa_0^2
\left[ 2C^{\mu\lambda\nu\kappa}_{;\lambda ; \kappa} - 
C^{\mu\lambda\nu\kappa}R_{\lambda \kappa}\right]\nonumber\\ = \
\kappa_0^2T^{\mu\nu}_{\rm matter}~.
\eeq
We are interested in the cosmology associated with these equations
of motion. For a Friedmann-Lema\^{i}tre-Roberston-Walker (FLRW) 
space-time, the Weyl tensor vanishes. Hence, the noncommutative 
geometry corrections to the Einstein equation, Eq.~(\ref{eq:EoM2}),
vanish. 

For scalar perturbations around a FLRW metric, in the conformal
Newtonian (also called longitudinal) gauge, the metric reads:
\beq\label{scal-pert} g_{\mu\nu} = {\rm diag} \left( \{a(t)\}^2 \left[
  -(1+\Psi(x)),\right.\right.\nonumber\\  \left.  (1-\Phi(x)),
  (1-\Phi\left(x)), (1-\Phi(x)) \right]\right)~,  \eeq
where $t$ is conformal time and $(x,y,z)$ are Cartesian space
coordinates; the scale factor is denoted by $a$.  The $\Phi, \Psi$ are
the gauge invariant {\sl Bardeen potentials}~\cite{Mukhanov:1990me};
the $\Psi$ is the analog of the Newtonian potential.

The $(0,0)$-component of Eq.~(\ref{eq:EoM2}) leads, for the metric
specified by Eq.~(\ref{scal-pert}), to the modified Friedmann equation:
\beq \label{eq:fried1}-3\left(\frac{\dot{a}}{a}\right)^2  -{\bf
  \nabla}^2\Phi\left(x\right) +3\left( \frac{\dot{a}}{a}\right)
\dot{\Phi}\left(x\right)\left(x\right)\nonumber\\ 
-\frac{\alpha_0\kappa_0^2 }{3a^2} {\bf \nabla}^4 \left[
  \Phi\left(x\right) + \Psi\left(x\right)\right]+ {\cal O}\left(
\Phi^2, \Psi^2, \dots\right)\nonumber\\
 = \kappa_0^2 T_{00}~; \eeq
over-dot denotes derivative {\sl w.r.t} conformal time $t$ and ${\bf
  \nabla}_i = \left( \partial_x, \partial_y, \partial_x\right)$.  

As expected, in an exactly homogeneous and isotropic space-time, {\sl
  i.e.},  $\Psi\left(x\right)=\Phi\left(x\right)=0$, the modified
Friedmann equation reduces to its standard form.  Naively, one may
have expected this result, since in a spatially homogeneous space-time
the spatial points are equivalent and hence any noncommutative effects
might be expected to vanish. This however is not the case here, since
the noncommutativity of the theory is incorporated in the internal
manifold $F$ and our space-time is a (smooth) commutative
four-dimensional manifold.  Despite this, noncommutative corrections
to the standard cosmological models do not occur at the level of an
FLRW background.  Notice that in the case of an FLRW model, one can
explicitly calculate the topological term $\int R^\star R^\star
\sqrt{g} {\rm d}^4 x$, appearing in Eq.~(\ref{eq:action1}), and show
that it is indeed nondynamical. 

Considering the scalar perturbations, the noncommutative geometry
corrections are in second and fourth order in spatial derivatives,
which can be neglected in most cosmological situations.

Of the remaining equations of motion, given in Eq.~(\ref{eq:EoM1}),
the most interesting ones are those coming from the off-diagonal
terms, namely
\beq -\partial_i\left[
\dot{\Phi}(x) + \frac{\dot{a}}{a}\Psi(x)\right] - \frac{\alpha_0
  \kappa_0^2}{6a^2} {\bf \nabla}^2\left[ \partial_i\left( \dot{\Psi}(x) +
\dot{\Phi}(x)\right)\right]\nonumber\\
 = \kappa_0^2 T_{0i}~, \eeq
and
\beq\label{eq:shear}
&& \frac{1}{2}\partial_i\partial_j \left[\Psi(x)-\Phi(x)\right] \nonumber \\
&& + \frac{\alpha_0
\kappa_0^2}{12a^2} \left[ 3\frac{\ddot{a}}{a} - 6\left( \frac{\dot{a}}{a}
\right)^2 - 3\partial^2_t + {\bf \nabla}^2 \right]\nonumber\\
&&\ \ \ \ \ \ \ \times \Big[
 \partial_i \partial_j \left( \Psi(x) + \Phi(x)\right) \Big]
\nonumber \\
&& =
\kappa_0^2 T_{ij}~~\mbox{with}~~i\neq j~.
\eeq
Equation (\ref{eq:shear}) is particularly interesting: it shows that
matter with zero anisotropic stresses no longer implies equality of
the Bardeen potentials ({\sl i.e.}, the condition $\Psi = \Phi$ does
not hold). Let us emphasise that in absence of noncommutative effects
({\sl i.e.}, for standard scalar perturbations in a FLRW background),
  the Bardeen potentials turn out to be equal, if shear-free matter
  fields are considered.

The above calculation can be also performed in the synchronous
gauge~\footnote{It corresponds to having only two nonzero perturbation
  variables; the other two being zero imply that the {\sl threading}
  of space-time into lines (corresponding to fixed space coordinates)
  consists of geodesics and the {\sl slicing} into hyper-surfaces
  (corresponding to fixed time) is orthogonal to them. There is a
  whole class of gauges with this property.}, for which the total
({\sl i.e.}, background + perturbed) metric can be written as:
\be g_{\mu\nu} = {\rm diag} \left( \{a(t)\}^2 \left[
  -1,\left(\delta_{ij} + h_{ij}\left(x\right)\right) \right]\right)~,
\ee
leading to the modified Friedmann equation:
\beq\label{eq:Friedmann}
&&-3\left(\frac{\dot{a}}{a}\right)^2 + \frac{1}{2}\left[
4\left( \frac{\dot{a}}{a}\right)\dot{h} + 2 \ddot{h}
- {\bf \nabla}^2 h + \nabla_i \nabla_j h^{ij}\right]
\nonumber \\
&& - \frac{\alpha_0\kappa_0^2}{6a^2} \left[
\partial_t^2 \left( {\bf \nabla}^2h - 3 \nabla_i
\nabla_j h^{ij} \right) 
\right.
\nonumber\\
&&
\ \ \ \ \ \ \ \ \ \ \ 
\left.
+ {\bf \nabla}^2\left(
\nabla_i \nabla_j h^{ij} \right) - {\bf \nabla}^4 h \right]
\nonumber \\
&& +{\cal O}\left( h^2 \right) \ \ =\ \  \kappa_{0}^2 T_{00}~,
\eeq
where $h \equiv h^i_i$ is the trace of $h_{ij}$.  

The remaining gauge is removed by choosing ${\bf \nabla}_i h^{ij}=0$,
for which Eq.~(\ref{eq:Friedmann}) reduces to:
\beq
-3\left(\frac{\dot{a}}{a}\right)^2 + 2\left(\frac{\dot{a}}{a}\right)
\dot{h} + \ddot{h} - \frac{1}{2}{\bf \nabla}^2 h 
\nonumber \\
-\frac{\alpha_0\kappa_0}{6a^2} {\bf \nabla}^2 \left[\partial_t^2 -{\bf
    \nabla}^2\right]h
+{\cal O}\left( h^2 \right) \ \ =\ \  \kappa_{0}^2 T_{00}~.  
\eeq
Thus, the traceless part of the perturbed metric $h_{ij}$, {\sl i.e.},
gravitational waves (which are in addition transverse), do not enter
into the Friedmann equation, even in the presence of noncommutative
geometry corrections. 

The remaining equations obtained from Eq.~(\ref{eq:EoM2}) are rather
involved, however for perturbations around a Minkowski background,
({\sl i.e.}, $a\left(t\right)=1$, $\dot{a}=0$) there is a significant
simplification due to the fact that
\be
C^{\mu\lambda\nu\kappa}R_{\lambda\kappa}\sim {\cal O}\left(
\left\{h_{ij}\right\}^2\right)~, \ee
where $\left\{h_{ij}\right\}$ indicates all terms that are first order
in the perturbation. For this situation, the transverse, traceless
part of $h_{ij}$ obey the following equations (where without loss of
generality we have taken $h_{ij}$ to be transverse to the $z$
direction, and we have again used the gauge condition ${\bf \nabla}_i
h^{ij}=0$):
\beq\label{eq:grav_wave1} \left[ 1 + \alpha_0\kappa_0^2 \left(
  -\partial_t^2 + \partial_z^2\right)\right] \left( - \partial_t^2 +
\partial_z^2\right) h_+ &=& 0~, \\
\label{eq:grav_wave2}
\left[ 1 + \alpha_0\kappa_0^2 \left( -\partial_t^2 + \partial_z^2\right)\right]
\left( - \partial_t^2 + \partial_z^2\right) h_\times &=& 0~,
\eeq
where $h_+$ and $h_\times$ are the two independent polarisations of the
gravitational waves, {\sl i.e.},
\be
 h_{ij} = \left(\begin{array}{ccc}
         h_+ & h_\times & 0 \\
         h_\times & -h_+ & 0 \\
         0 & 0 & 0 \end{array}\right)~.
\ee
The right-hand side of
Eqs.~(\ref{eq:grav_wave1}),~(\ref{eq:grav_wave2}) vanish because we
are considering gravitational waves propagating against a Minkowski
background, for which $T_{\mu\nu}=0$. 

It is clear from Eqs.~(\ref{eq:grav_wave1}),~(\ref{eq:grav_wave2})
that the solutions to the General Relativistic equation for the
components of the perturbations (produced here by setting $\alpha_0 =
0$) remain solutions, {\sl i.e.}, one finds that perturbations
satisfying,
\be \left( -\partial_t^2 + \partial_z^2\right) h_+ = 0~ \ \ \ {\rm
  and}\ \ \ \left( -\partial_t^2 + \partial_z^2\right) h_\times = 0~,
\ee are solutions to the equations of motion
(Eq.~(\ref{eq:EoM2})). 

Thus, the propagation of standard gravitational
waves is unaffected by the presence of noncommutative geometry effects
(at least for gravitational waves propagating in Minkowski
space-time). However, there are additional solutions to
Eqs.~(\ref{eq:grav_wave1}),~(\ref{eq:grav_wave2}), which
correspond to {\sl gravitational radiation}, that are not present in
standard General Relativity. A detailed investigation of this
phenomenon is performed in Ref.~\cite{us_to_do}.

In order for the corrections to Einstein's equations to be apparent at
leading order, ({\sl i.e.}, at the level of the background),
we need to consider anisotropic models. As an example,
we calculate the modified Friedmann equation for the Bianchi type-V
model, for which the space-time metric, in Cartesian coordinates,
reads
\be g_{\mu\nu} = {\rm diag} \left[ -1,\{a_1(t)\}^2e^{-2nz} ,
 \{a_2(t)\}^2e^{-2nz}, \{a_3(t)\}^2 \right]~, \ee
where $a(t)$, $b(t)$ and $c(t)$ are, in general, arbitrary functions
and $n$ is an integer. 

Defining
$A_i\left(t\right) = {\rm ln} a_i\left(t\right)$  with $i=1,2,3$,
the modified Friedmann equation reads:
\beq\label{eq:Friedmann_BV} \kappa_0^2 T_{00}=&&\nonumber\\
 - \dot{A}_3\left(
\dot{A}_1+\dot{A}_2\right) -n^2 e^{-2A_3} \left( \dot{A}_1
\dot{A}_2-3\right)&& \nonumber \\
 +\frac{8\alpha_0\kappa_0^2 n^2}{3} e^{-2A_3} \left[
  5\left(\dot{A}_1\right)^2 + 5\left(\dot{A}_2\right)^2 -
  \left(\dot{A}_3\right)^2\right.&&\nonumber
\\
\left. - \dot{A}_1\dot{A}_2 - \dot{A}_2\dot{A}_3
  -\dot{A}_3\dot{A}_1 - \ddot{A}_1 - \ddot{A}_2 - \ddot{A}_3 + 3
  \right]&& \nonumber \\
- \frac{4\alpha_0\kappa_0^2}{3} \sum_i \Biggl\{
\dot{A}_1\dot{A}_2\dot{A}_3 \dot{A}_i&&\nonumber\\
 + \dot{A}_i \dot{A}_{i+1} \left(
\left( \dot{A}_i - \dot{A}_{i+1}\right)^2 -
\dot{A}_i\dot{A}_{i+1}\right)&& \nonumber \\
 + \left( \ddot{A}_i + \left( \dot{A}_i\right)^2\right)\left[
  -\ddot{A}_i - \left( \dot{A}_i\right)^2 + \frac{1}{2}\left(
  \ddot{A}_{i+1} + \ddot{A}_{i+2} \right)\right.&&\nonumber\\
\left. + \frac{1}{2}\left(
  \left(\dot{A}_{i+1}\right)^2 + \left( \dot{A}_{i+2}\right)^2 \right)
  \right]&& \nonumber \\ 
+ \left[ \dddot{A}_i + 3 \dot{A}_i \ddot{A}_i -\left( \ddot{A}_i +
  \left( \dot{A}_i\right)^2 \right)\left( \dot{A}_i - \dot{A}_{i+1} -
  \dot{A}_{i+2} \right)\right]&&\nonumber\\
\times\left[ 2\dot{A}_i
  -\dot{A}_{i+1}-\dot{A}_{i+2} \right]\Biggr\}~ \eeq
where all indices are understood to be taken modulo $3$ and the $t$
dependence of the terms has been suppressed.  Clearly, all terms
containing $\alpha_0$, in Eq.~(\ref{eq:Friedmann_BV}) above, are the
modifications to the standard result.  By studying the case of the
Bianchi type~V model, we can immediately identify the noncommutative
geometry effects in other cases of cosmological model. More precisely,
Eq.~(\ref{eq:Friedmann_BV}) reduces to:
\begin{itemize}
\item
Bianchi type-I for
$n=0$;
\item
 FLRW for $a(t) = b(t) = c(t)$ and $n=0$; 
\item
Kasner metric~\footnote{A sub-class of the Bianchi type-I metrics.}
for $a(t) = t^A$, $b(t)=t^B$, $c(t)=t^C$ and $n=0$, where $A$, $B$ and
$C$ are constants.
\end{itemize}

For the Bianchi type-V metric, with 
\be a(t) =
t^{\tilde{a}_1}~~,~~b(t)=t^{\tilde{a}_2}~~,~~c(t)=t^{\tilde{a}_3}~~,
\nonumber \ee 
where $\tilde{a}_i$ are constants as in the Kasner metric but $n
\neq 0$, the modified Friedmann equation becomes:
\beq \kappa_0^2 T_{00} = - \tilde{a}_3 \left(
\tilde{a}_1+\tilde{a}_2\right) t^{-2} - n^2 t^{-2\left(
  \tilde{a}_3+1\right)} \left( \tilde{a}_1 \tilde{a}_2 -3\right)&&
\nonumber \\ 
+ \frac{8\alpha_0\kappa_0^2n^2}{3} t^{-2\left(
  \tilde{a}_3+1\right)} \left[ 5\left( \tilde{a}_1\right)^2 + 5\left(
  \tilde{a}_2\right)^2 
\right.&&
\nonumber\\
\left.
- \left( \tilde{a}_3\right)^2 -
  \tilde{a}_1\tilde{a}_2 - \tilde{a}_2 \tilde{a}_3 -
  \tilde{a}_3\tilde{a}_1 + \tilde{a}_1 + \tilde{a}_2 + \tilde{a}_3 +3
  \right]&& \nonumber \\
 -\frac{4\alpha_0\kappa_0^2}{3}t^{-4} \sum_i \tilde{a}_i \Biggl\{
\tilde{a}_1\tilde{a}_2\tilde{a}_3&&\nonumber\\
 + \tilde{a}_{i+1}\left( \left(
\tilde{a}_i - \tilde{a}_{i+1}\right)^2 - \tilde{a}_i
\tilde{a}_{i+1}\right) \nonumber&& \\
+ \left( \tilde{a}_i - 1\right) \left[ \frac{1}{2}
  \tilde{a}_{i+1}\left( \tilde{a}_{i+1} -1\right) + \frac{1}{2}
  \tilde{a}_{i+2}\left( \tilde{a}_{i+2} -1\right)\right.&&\nonumber\\
\left. - \tilde{a}_i \left(
  \tilde{a}_i -1\right) \right] \nonumber &&\\ 
+ \left[ \left( \left(\tilde{a}_i\right)^2 +2 \right) -3\tilde{a}_i
\right.
  &&\nonumber\\ \left.+ ( 1-\tilde{a}_i) ( \tilde{a}_i -
  \tilde{a}_{i+1} - \tilde{a}_{i+2} ) \right] \left[
  2\tilde{a}_i - \tilde{a}_{i+1} -\tilde{a}_{i+2}\right] \Biggr\}~.&&
\eeq
Since the term in braces occurs at a higher order than the terms
coming from the Einstein-Hilbert action (at least for $\tilde{a}_3
<1$), it becomes negligible at late times.

For the Kasner metric we know that $n=0$ and hence the only
correction to the standard Friedmann equation is the term in
braces. However, for the inhomogeneous case ($n\neq 0$) there is an
additional term that occurs at the same order as the inhomogeneous
part of the standard Friedmann equation, {\sl i.e.}, at order
$t^{-2\left( \tilde{a}_3+1\right)}$.

More generally, from Eq.~(\ref{eq:Friedmann_BV}) the correction terms
come in two types. The first one contains the terms in braces in
Eq.~(\ref{eq:Friedmann_BV}), which are fourth order in time
derivatives. Hence for the slowly varying functions, usually used in
cosmology, they can be taken to be small corrections. The second type,
which is the third term in Eq.~(\ref{eq:Friedmann_BV}), occurs at the
same order as the standard Einstein-Hilbert terms. However, it is
proportional to $n^2$ and hence vanishes for homogeneous versions of
Bianchi type-V. Thus, although anisotropic cosmologies do contain
corrections due to the additional NCG terms in the action, they are
typically of higher order.  Inhomogeneous models do contain correction
terms that appear on the same footing as the original (commutative)
terms.

\subsection{\bf Nonminimal coupling of the Higgs  field to curvature}

Up to now, we have neglected the non-minimal coupling of the Higgs
field to the curvature. This is likely to be a good approximation for
late time cosmology, since we expect the Higgs field to be very
small. However, at energies approaching the Higgs scale this
additional term needs to be included.  From Eq.~(\ref{eq:EoM1}) it is
immediately apparent that for $|{\bf H}|\neq 0$, the effects of the
NCG corrections to Einstein's equations are enhanced.  In particular,
for $|{\bf H}|\rightarrow \sqrt{6}/\kappa_0$ the correction term
entirely dominates, provided the Weyl curvature term is nonzero, and
the equations of motion tend to
\be 2 C^{\mu\lambda\nu\kappa}_{;\lambda,;\kappa} -
C^{\mu\lambda\nu\kappa} R_{\lambda\kappa} =
-\frac{1}{\alpha_0} T^{\mu\nu}_{\rm matter}~, \ee 
which is precisely the equations of motion for conformal
gravity~\cite{mannheim}, albeit with a modified gravitational
constant.

As we have previously shown, the corrections to Einstein's equations
are present only in inhomogeneous and anisotropic space-times. For
$|{\bf H}|\neq 0$ however, there are corrections even for background
cosmologies. To understand the effects of these corrections it is
sufficient to neglect the conformal term in Eq.~(\ref{eq:EoM1}),
{\sl i.e.}, setting $\alpha_0 = 0$. In this case, the equations of motion
become:
\be R^{\mu\nu} - \frac{1}{2}g^{\mu\nu}R =
\kappa_0^2\left[\frac{1}{1-\kappa_0^2 |{\bf H}|^2/6}\right] T^{\mu\nu}_{\rm
  matter}~. \ee 
Hence, the effect of a nonzero $|{\bf H}|$ field is to create an
effective gravitational constant.

An alternative view point is to consider the effect of this term on
the equations of motion for the Higgs field in some, constant,
gravitational field. The action for the pure Higgs fields reads~\cite{ccm}
\be\label{eq:action_Higgs} {\cal L}_{|{\bf H}|} = -\frac{R}{12}|{\bf
  H}|^2 + \frac{1}{2} |D^\alpha {\bf H} | | D ^\beta {\bf H} |
g_{\alpha\beta} - \mu_0 |{\bf H}|^2 + \lambda_0|{\bf H}|^4~; \ee
$D^\alpha$ is the covariant derivative. Thus, for constant curvature,
the self interaction of the Higgs field is increased, namely
\be -\mu_0 |{\bf H}|^2 \rightarrow -\left( \mu_0 + \frac{R}{12}\right)
|{\bf H}|^2~.  \ee
Hence, for static geometries, the nonminimal coupling of the Higgs
field to to the curvature increases the Higgs mass. This has
potential consequences both for terrestrial experiments and for late
time cosmology, since the curvature of an asymptotically de Sitter
universe would increase the effective mass of the Higgs field,
although in both cases the effect is likely to be minimal. 

\section{Discussion}
After having discussed some cosmological consequences of the
noncommutative geometry spectral action, let us briefly mention some
links to dilatonic gravity and chameleon cosmology, in the presence of
the nonminimal coupling of the Higgs field to the background
geometry.

Redefine the Higgs field ${\bf H}$ by
\be \tilde{\phi} = -\ln \left( |{\bf H}|/(2\sqrt{3})\right)~,
\nonumber
\ee
and thus rewrite Eq.~(\ref{eq:action_Higgs}) in the form of
four-dimensional dilatonic gravity as
\beq {\cal L}_{\tilde{\phi}} = e^{-2\tilde{\phi}} \left[ -R +
  6D^\alpha \tilde{\phi}D^\beta \tilde{\phi} g_{\alpha\beta}\right.
  \nonumber\\ \left.  - 12\left( \mu_0 -12\lambda_0
  e^{-2\tilde{\phi}}\right) \right], \eeq
providing a link to compactified string models. 

In chameleon models~\cite{chameleon}, a scalar field is taken to have
a nonminimal coupling to the standard matter content (thus evading
solar system tests of General Relativity).  In the NCG spectral action
studied here, we have a scalar field (the Higgs) that has a nonzero
coupling to the background geometry. If we are in a regime where the
equations of motion are well approximated by Einstein's equations,
then the background geometry will be given (approximately) by the
standard matter, making the mass and dynamics of the Higgs field
explicitly dependent of the local matter content. A more detailed
study of this link to chameleon models is left as a future
work~\cite{us2}.

The noncommutative geometry spectral action gives an elegant
mathematical formulation of the Standard Model of elementary particle
physics, compatible with all known phenomenology of the Standard
Model. In addition, it provides a natural set-up to study early
universe cosmology~\footnote{After completion of this work,
  cosmological studies have been also performed in
  Refs.~\cite{infl,mp}.}.

\section{Acknowledgements}
It is a pleasure to thank Ali Chamseddine, Alain Connes and John
Barrett for enlightening discussions.  The work of M.~S. is partially
supported by the European Union through the Marie Curie Research and
Training Network {\sl UniverseNet} (MRTN-CT-2006-035863).

\end{document}